\documentclass[preprint]{revtex4-1}
\usepackage[english]{babel}
\usepackage{amsmath}
\usepackage{amssymb}
\usepackage[final]{graphicx}
\usepackage{upgreek}
\usepackage[utf8]{inputenc}

\newcommand{\distance}{5mm}			
\usepackage{color}

\begin{document}
\title{Discrete relaxation of exciton-polaritons in an inhomogeneous potential}
\author{T. Michalsky}
\author{H. Franke}
\author{C. Sturm}
\author{M. Grundmann}
\author{R. Schmidt-Grund}
\affiliation{ 
Universit\"{a}t Leipzig, Linn\'{e}stra{\ss}e 5, D-04103 Leipzig,
Germany
}%
\date{\today}

\begin{abstract}
\noindent We present indications, that the wave function-stiffness condition during energy-relaxation as observed in single-phase state quantum systems manifests also in a single particle ensemble. This is demonstrated for exciton-polaritons in the strong coupling regime in a ZnO-based microcavity at $T=10\ \mathrm{K}$ for non-resonant excitation. It is well known that the pump-induced spatially inhomogeneous background potential leads to nearly equally spaced energy levels in the \textit{k}-space distribution for propagating polariton Bose-Einstein condensates. Surprisingly this particular pattern is also observable for uncondensed exciton-polaritons.

\end{abstract}

\pacs{xyz}

\maketitle
\section{Introduction}
\noindent\hspace*{\distance}
In microcavities (MCs) exciton-polaritons (polaritons) are composite particles with partly photonic and excitonic properties. Their bosonic character allows them to undergo the phase transition to a macroscopically coherent state known as a dynamic Bose-Einstein condensate (BEC) which was first experimentally shown in 2006 in CdTe-based MCs \cite{kasp} at liquid helium temperatures. Since then BECs in MCs have evolved to be a model system for research in many topics of modern physics, like superfluidity~\cite{amo}, vortex-generation~\cite{lagou} and ballistic transport phenomena~\cite{wertz} leading to novel devices like all-optical polariton switches~\cite{li} and transistors~\cite{anton}. In this context often a varying background potential is used to accelerate the condensed particles. Relaxation processes connected to this background potential have been observed by several groups~\cite{wertz, guillet, christop, krizh09, helena}. Wouters \textit{et al.}~\cite{wouters2} developed a description of the occuring phenomena in terms of energy conservation and scattering properties. In this work, we present experimental results suggesting that this model can also be applied to a certain extent to an ensemble of uncondensed polaritons. \\
\section{Experimental Methods}
\noindent\hspace*{\distance}
The investigated ZnO-based  MC is the same as has been described in Ref.~\cite{helena}. The MC contains a wedge-shaped cavity layer giving access to a wide range of detuning values $\mathrm{\Delta}$ between the uncoupled cavity-photon mode and the excitonic transition energy.
We used a frequency tripled and mode\-locked Ti:Sa laser at a wavelength of 266 nm for non resonant excitation. The pulse width was 2 ps at a repetition rate of 76 MHz. 
The laser beam was focused on an area of  about 5 $\upmu \mathrm{m^{2}}$ through an UV objective with a magnification of 50 and a numerical aperture (NA) of 0.4 corresponding to a detectable angular range of $\pm 23^\circ$. The detection of the light emitted from the sample was realized in a confocal configuration. The lens setup allows for the imaging of the momentum (\textit{k}-) as well as of the real space. 
In order to investigate relaxation processes which take place in the region where the pump-induced background potential is strongly inhomogeneous, a pinhole (PH) was put in an intermediate image plane so that only the emission from the central excitation area of about 3~$\upmu \mathrm{m^2}$ was detected.
If the PH is used the lower polariton branch (LPB) emission is spectrally broadened and smeared out. This is caused by diffraction at the PH as follows from the Heisenberg principle: $\mathrm{\Delta} x\mathrm{\Delta} k\geq 0.5$. Furthermore the detectability of high-$k_{||}$ emission is suppressed as the cryostat window introduces spherical aberrations which strongly increase with the emission angle causing these rays to be blocked at the PH. \\
\section{Experimental Results and Discussion}
\begin{figure*}[htb]
	\centering
  \includegraphics[width=1\textwidth, angle=0]{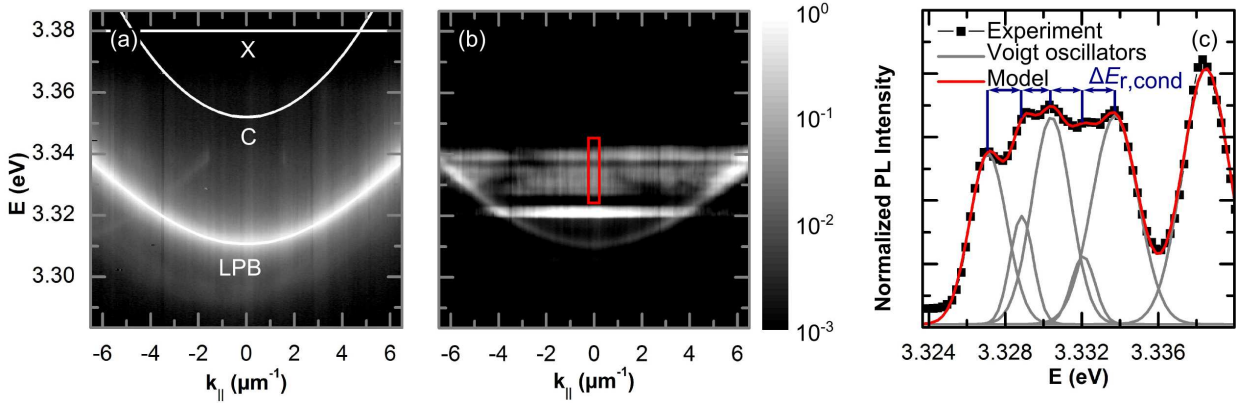}
	\caption{Energy resolved \textit{k}-space image of the LPB for $\mathrm{\Delta}=-30\ \mathrm{meV}$ at $T=10\ \mathrm{K}$ (a) below ($0.04\ \mathrm{P_{th}}$) and (b) above ($1.6\ \mathrm{P_{th}}$)  the condensation threshold. One has to note that there is a pale ghost image of the LPB shiftes by about $8\ \upmu \mathrm{m^{-1}}$ to negative $k_{||}$ values produced by the monochromator in (a). The modeled uncoupled exciton and the cavity-photon dispersion are marked with X and C, respectively. (c) shows the PL spectrum at $k_{||} = 0\ \upmu \mathrm{m^{-1}}$ from the red marked region in (b). $\mathrm{\Delta} E_{\mathrm{r,cond}}$ marks the energy spacing between the dispersionless condensate states.
In order to specify the spectral position of these states the measured data (black symbols) were fitted with Voigt oscillators (gray lines).}
	\label{fig:woph}
\end{figure*}

\noindent\hspace*{\distance}
The following results are taken from a sample position corresponding to $\mathrm{\Delta}=-30 \ \mathrm{meV}$ ($\sim 0.6 \times$ the coupling constant of $V=50\ \mathrm{meV}$) at $T=10\ \mathrm{K}$. Figure~$\ref{fig:woph}$ shows the polariton emission below (a) and above (b) the condensation threshold without PH and includes the calculated dispersion relations of the uncoupled exciton (X) and cavity-photon mode (C)~\cite{chris}. In the uncondensed case the lower polariton branch (LPB) with a homogeneous broadening of $\gamma_{\mathrm{uncond}}= (2.8\pm 0.2)\ \mathrm{meV}$ is clearly visible without any recognizable further features.
Above the condensation threshold  (Fig. $\ref{fig:woph}$ (b) and (c)) several dispersionless states appear bounded by the dispersion relation of the uncondensed polaritons. These states are roughly equally spaced in energy (see Fig. \ref{fig:woph} (c)) with a distance of  $\mathrm{\Delta} E_{\mathrm{r,cond}}= (1.7\pm 0.3)\ \mathrm{meV}$ and a homogeneous broadening of $\gamma_{\mathrm{cond}}= (1.5\pm 0.2)\ \mathrm{meV}$. Thus one can conclude that the condensate lifetime is roughly twice as high as for the uncondensed polaritons. The relative occupation of the condensate states changes with in-plane momentum and does not show a regular pattern.
Please note that the spectral broadening of the condensate states shown in Fig.~\ref{fig:woph}~(b) and (c) is increased additionally to the lifetime broadening by temporal potential fluctuations due to the short Ti:Sa pump-laser pulses \cite{sanvitto}. The homogeneous broadening of the condensate emission was determined via fitting Voigt-oscillators to PL spectra obtained from experiments~\cite{helena,diss_helena} with a diode laser with a pulse duration of $500\ \mathrm{ps}$ acting as quasi continuous excitation and yielding spectrally narrower condensate states.
The lowest energy condensate state at $3.320\ \mathrm{eV}$ is excluded in Fig. \ref{fig:woph} (c) because of its very strong inhomogeneous broadening. This is caused by the relatively long effective lifetime of this condensate state whose energy follows the decreasing background potential as time resolved measurements (not shown here) reveal.  

\begin{figure}[htb]
	\centering
  \includegraphics[width=0.50\textwidth, angle=0]{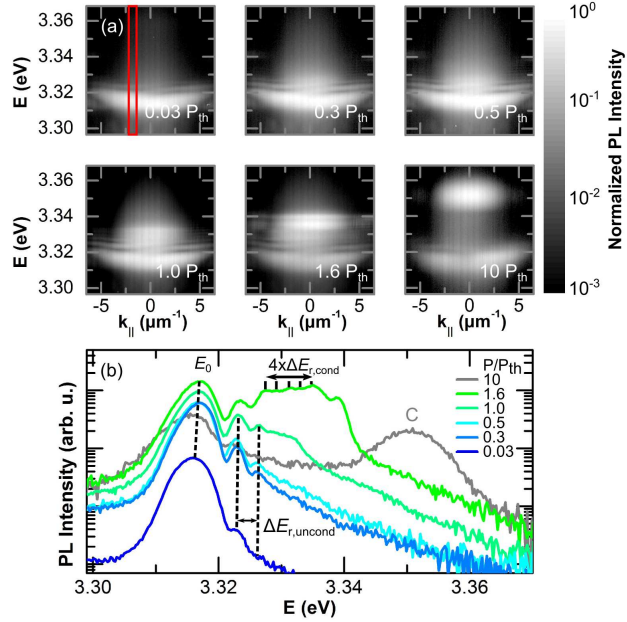}
	\caption{(a) shows the normalized energy resolved \textit{k}-space images for different excitation densities. The red bar in the first image marks the $k_{||}$ channel the spectra in (b) are taken from. Note that the intensity scale is logarithmic. In (b) $E_{0}$ represents the polariton emission of lowest energy, $\mathrm{\Delta}E_{\mathrm{r,uncond}}$ marks the energy spacing between two emission states which correspond to blueshifted polaritons and C represents the energy of the uncoupled cavity mode.}
	\label{fig:images}
\end{figure}

\begin{figure}[htb]
	\centering
  \includegraphics[width=0.50\textwidth, angle=0]{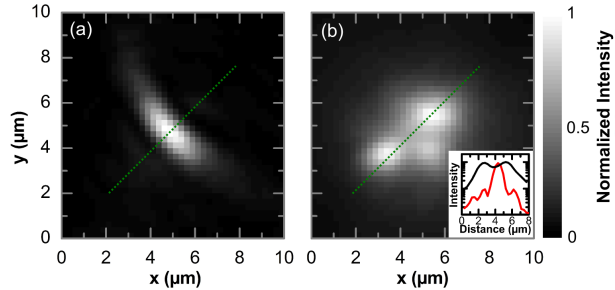}
	\caption{The real space image of the laser spot on the sample surface is shown in (a) whereas  the LPB emission below the condensation threshold is shown in (b). The inset in (b) shows the normalized intensity profile of the laser spot (red) and the polariton emission (black) along the dashed green line.}
	\label{fig:spot}
\end{figure}

Regarding the \textit{k}-space pattern for increasing excitation power, the situation changes when only photons are detected which were emitted from the central excitation area where the pump-induced background potential is strongly inhomogeneous. This is shown in Fig. \ref{fig:images}, where one can clearly see that in addition to the dominating LPB two further polariton branches become visible at higher energies even far below the condensation threshold. It is important to note that for increasing excitation density the distance in energy of $\mathrm{\Delta}E_{\mathrm{r,uncond}}= (3.3\pm 0.1)\ \mathrm{meV}$ between these additional branches stays constant as demonstrated more clearly by the  $k_{||} = -2\ \upmu\mathrm{m^{-1}}$ spectra in Figure \ref{fig:images} (b). As the excitation density increases the LPB branches are shifted towards higher energies caused by the repulsive Coloumb interaction between the polaritons according to their excitonic parts and the interaction with hot carriers \cite{ciuti}. One should be aware of the fact that these additional branches are not symmetrical to $k_{||} = 0\ \upmu\mathrm{m^{-1}}$ because the irregular bar-like laser spot profile, a consequence of the birefringent frequency tripling crystals,  destroys rotational symmetry, as shown in Fig.~\ref{fig:spot}~(a). At high excitation densities ($10\ \mathrm{P_{th}}$ in Fig. \ref{fig:images} (a) and \ref{fig:images} (b)) the bare cavity-photon mode becomes visible as the particle density reaches the Mott transition. This is accompanied by the disappearance of condensate emission.
As mentioned above, at the condensation threshold new dispersionless states appear in the energy resolved \textit{k}-space images being roughly equally spaced in energy. They belong to the coherent emission of condensates \cite{martin} being expelled from the central excitation area \cite{wertz, wouters1}. 
The appearance of these states in the condensed phase has been observed also by other groups~\cite{wertz, guillet, christop, krizh09} and was explained in \cite{wouters2, wouters1} as a consequence of the spatially inhomogeneous background potential induced by the optical pumping.
The roughly equal energy distances between these states were attributed to the dynamical balance of in- and outscattering rates for bosonic states of different energy \cite{wouters2}: 
\begin{equation}
\Gamma_{in}=\Gamma_{out}
\label{eq:1}
\end{equation} 
Following the ansatz of Wouters \textit{et al.} \cite{wouters2} the inscattering rate $\Gamma_{in}$ for a polaritonic (or bosonic) state is proportional to the energy distance to the initial state of the scattering process:
\begin{equation}
\Gamma_{in}=\kappa\times\mathrm{\Delta} E_{r}
\label{eq:2}
\end{equation}
Here the scattering constant $\kappa$ has a unit of an inverse action, for simplicity not being normalized to a unit density as in the original paper of Wouters \textit{et al.}~\cite{wouters1}.
The outscattering rate $\Gamma_{out}$ of a state is limited by its radiative lifetime and is therefore assumed to be proportional to the homogeneous spectral broadening $\gamma$:
\begin{equation}
\Gamma_{out}\simeq\frac{\gamma}{\hbar}
\end{equation}
This means that for states with a doubled outscattering rate ($\propto$\ doubled spectral broadening) the energetical distance $\mathrm{\Delta} E_{r}$ to the next state has to be twice as high to reach a compensation of in- and outscattering rates, as can be seen from equations (\ref{eq:1}) and (\ref{eq:2}). Exactly this is the case in our sample when we compare the condensate emission with the emission from the uncondensed polariton ensemble. From our experimental results we can deduce the bosonic scattering constant to be $\kappa = (0.8\pm 0.4)\ \hbar^{-1}$. This result, namely that the spectral broadening of a condensate state is roughly the same as the energy spacing between relaxing condensate states was also observed in  \cite{guillet, christop, wertz}.
The consistent applicability of this simple model to both, the condensed \textit{and} the uncondensed phase, leads to the conclusion  
that in a varying background potential discrete relaxation processes take place in both phases which can be described with one scattering constant $\kappa$ being independent of the particle density within the observed range of more than two orders of magnitude.

The relaxation in the potential landscape is accompanied with an effective acceleration towards the peripheral region of the excitation spot where the local potential energy is smaller. To show this, we recorded the laser spot intensity distribution on and the emission of uncondensed polaritons from the sample surface (Fig. \ref{fig:spot}). It can be clearly seen that the main emission of uncondensed polaritons is coming from the spatial region where the laser light intensity is smaller.
Here the uncondensed polaritons could be trapped in the area of the first minimum of the excitation laser spot (see inset of Fig.~\ref{fig:spot} (b)), providing another explanation for the appearance of the observed discrete polariton states, namely the quantization in a potential trap created by first diffraction minimum of the pump laser spot. But this seems not reasonable as both, the uncondensed and the condensed polaritons should feel more or less the same trapping potential resulting in equal energy spacings for the trapped uncondensed and condensed polaritons ($\Delta E_{\mathrm{r,uncond}}\overset{!}{=}\Delta E_{\mathrm{r,cond}}$), which is obviously not the case.       
\\
\section{Conclusion} 
\noindent\hspace*{\distance}
We have shown experimental results giving hints that the model \cite{wouters2} of relaxation processes describing condensed polaritons in a spatially varying potential landscape is also valid for the uncondensed phase. This assumption is supported by real and \textit{k}-space images revealing discrete polariton energy levels of constant (energy-) spacing in \textit{k}-space and an accumulation of polaritons in the peripheral region of the pump-induced potential hill shown by real space emission imaging. From our experimental results we could derive a value for the bosonic scattering constant in ZnO MC which is close to $\hbar^{-1}$.\\
\section{Acknowledgement}
\noindent\hspace*{\distance}
We thank H. Hochmuth and M. Lorenz for their support during the growth of the sample. Funding by DFG within Gru1011/20-2 and FOR1616 (SCHM2710/2-1) is acknowledged.

\bibliographystyle{apsrev4-1}

\providecommand{\noopsort}[1]{}\providecommand{\singleletter}[1]{#1}%

\end{document}